\documentclass[journal]{IEEEtran}
\usepackage{amsmath}
\usepackage{epsfig}
\usepackage{color}
\usepackage{xcolor}
\usepackage{amssymb}
\usepackage{flushend}
\usepackage{cite}
\usepackage[numbers,compress]{natbib}
\usepackage{subfigure}
\ifCLASSINFOpdf
\else
\fi
\begin{document}
%
\title{A Supervised Speech enhancement Approach with Residual Noise Control for Voice Communication}
%
%
%

\author{Andong Li,
        Chengshi~Zheng,~\IEEEmembership{Senior Member,~IEEE,}
        and~Xiaodong~Li
\thanks{The authors are with the Key Laboratory of Noise and Vibration Research, Institute of Acoustics, Chinese Academy of Sciences,
Beijing, 100190, China, and also with University of Chinese Academy of Sciences, Beijing, 100049, China (email: {cszheng}@mail.ioa.ac.cn).}
\thanks{This work was supported by NSFC
(National Science Fund of China) under Grant No. 61571435, Grant No. 61801468 and Grant No. 11974086.}
\thanks{Manuscript received October XX, 2019; revised December XX, XX.}}

%
%

\markboth{Journal of \LaTeX\ Class Files,~Vol.~XX, No.~XX, December~XXXX}%
{Shell \MakeLowercase{\textit{et al.}}: Bare Demo of IEEEtran.cls for Journals}
%



\maketitle

\begin{abstract}
For voice communication, it is important to extract the speech from its noisy version without introducing unnaturally artificial noise. By studying the subband mean-squared error (MSE) of the speech for unsupervised speech enhancement approaches and revealing its relationship with the existing loss function for supervised approaches, this paper derives a generalized loss function, when taking the residual noise control into account, for supervised approaches. Our generalized loss function contains the well-known MSE loss function and many other often-used loss functions as special cases. Compared with traditional loss functions, our generalized loss function is more flexible to make a good trade-off between speech distortion and noise reduction. This is because a group of well-studied noise shaping schemes can be introduced to control residual noise for practical applications. Objective and subjective test results verify the importance of residual noise control for the supervised speech enhancement approach.
\end{abstract}

\begin{IEEEkeywords}
Generalized loss function, residual noise control, noise shaping, speech distortion, deep learning.
\end{IEEEkeywords}

%
\IEEEpeerreviewmaketitle

\section{Introduction}
%
%
%
%
\IEEEPARstart{S}{peech} enhancement plays an important role in noisy environments for many applications, such as speech communication, speech interaction and speech translation. Numerous researchers have done lots of efforts on separating the speech from its noisy version and various approaches have already been proposed in the last five decades. Conventional approaches include spectral subtraction~{\cite{boll1979suppression}}, statistical method~{\cite{ephraim1984speech, ephraim1985speech}} and subspace-based method~{\cite{ephraim1995signal}}, which has proved to be valid when the additive noise is stationary or quasi-stationary. However, their performance often suffers from heavy degradation under non-stationary and low signal-to-noise ratio (SNR) conditions.

While moving head with deep learning, the supervised approaches gradually show their powerful capability on suppressing both stationary and highly non-stationary noise signals, which is mainly because of highly nonlinear mapping ability of deep neural networks (DNN)~{\cite{wang2014training}},~{\cite{xu2013experimental}}. In DNN-based algorithms, minimum mean-squared error (MSE) is often adopted as a loss criterion to update the weights of the network. Nevertheless, usage of this criterion directly may suffer from some problems. First, although MSE is the most often used criterion, it is not so relevant with speech perception~{\cite{hu2003perceptually, loizou2010reasons}}. Second, global MSE optimization usually obtains an over-smoothing estimation which omits some important detailed information. To solve these problems, many new criteria, that consider speech perception, have been proposed in most recent years~{\cite{martin2018deep, liu2017perceptually, kolbaek2018monaural, venkataramani2017adaptive}}. The first one is to use perceptually weighted MSE functions, which are proposed to weight the loss in different time-frequency (T-F) regions~{\cite{shivakumar2016perception, liu2017perceptually}}. The second one is to use objective metrics as loss functions, for examples, perceptual evaluation speech quality (PESQ)~{\cite{recommendation2001perceptual}}, short-time objective intelligibility (STOI)~{\cite{taal2010short}} and scale-invariant speech distortion ratio (SI-SDR)~{\cite{le2019sdr}} have been adopted as loss functions. In~{\cite{xu2019components}}, speech distortion and residual noise are considered separately in the loss function, which is called components loss (CL).





Note that all the above mentioned loss functions aim at suppressing noise as much as possible at noise-only segments. In other words, at noise-only segments, the amount of noise reduction is expected to be a positive infinite value. As we know, this aim could not be achieved in most cases for many reasons. First, the noise is often stochastic, and thus it is inevitable that the estimation accuracy is often constrained by a limited number of available observations~{\cite{zheng2011two, zheng2012statistical}}. Second, there are a great variety of noise signals, so that a DNN model cannot be expected to distinguish all of them correctly from the speech in each T-F unit. Therefore, when the noise cannot be suppressed totally as expected, some unnatural residual noise may degrade speech quality a lot~{\cite{gelderblom2018subjective}}, which needs to be considered carefully. In this paper, we derive a generalized loss function by introducing multiple manual parameters to flexibly make a balance between speech distortion and noise attenuation. More specifically, the residual noise control is introduced for voice communication~{\cite{gustafsson1998novel, braun2015residual}}. By theoretical derivations, MSE and other often-used loss functions can be included in the proposed generalized loss function.

The remainder of the paper is structured as follows. Section II formulates the problem. Section III derives the generalized loss in detail and introduces used network architecture. Section IV is the experimental settings. Results and analysis are given in Section V. Section VI presents some conclusions.

%




\section{Problem Formulation}
In the time domain, the noisy signal can be modelled as
\begin{equation}
\label{eq1}
\begin{array}{l}
 x\left( n \right) = s\left( n \right) + d\left( n \right),
 \end{array}
\end{equation}
where $s\left( n \right)$ is the clean speech and $d(n)$ is the additive noise. In the frequency domain, (\ref{eq1}) can be written as
\begin{equation}
\label{eq2}
\begin{array}{l}
 X_l\left( k \right) = S_l\left( k \right) + D_l\left( k \right),
 \end{array}
\end{equation}
where $X_l\left( k \right)$, $S_l\left( k \right)$, and $D_l\left( k \right)$ are, respectively, discrete Fourier transforms (DFT) of $x(n)$, $s(n)$, and $d(n)$ with the frame index $l$ and the frequency
bin $k$.

For practical applications, we only have the time-domain noisy signal $x(n)$ or its frequency-domain version
$X_l(k)$, the problem becomes how to estimate $s(n)$ or $S_l(k)$ from its noisy signal. It is common to
use Minimum MSE (MMSE) as a criterion in unsupervised speech enhancement approaches. Before introducing
MMSE, we first define the square error as
\begin{equation}
\label{eq3}
J_x \left[ {M_l\left( {k} \right)} \right] = \left| {f\left( {{S_l\left( {k} \right)}} \right) - g\left( {S_l\left( {k} \right), {D_l\left( {k} \right), M_l\left( {k} \right)} } \right)} \right|^2,
\end{equation}
where $M_l\left( {k} \right)$ is a nonlinear spectral gain function, $f(a)$ is a function with a variable $a$, and $g(a,b,c)$ is a function with three variables $a$, $b$, and $c$.
When $f(a) = |a|$ and $g(a,b,c) = |(a+b)c|$, $\mathop {\min }\limits_{M_l \left( k \right)} E\left\{ J_x \left[ {M_l \left( k \right)} \right]\right\}$ results in MMSE
spectral amplitude estimator in {\cite{ephraim1984speech}}, where $E\left\{\bullet\right\}$ is the expectation operator. When $f(a) = \log(|a|)$ and $g(a,b,c) = \log(|(a+b)c|)$, $\mathop {\min }\limits_{M_l \left( k \right)} E \left\{J_x \left[ {M_l \left( k \right)} \right] \right\}$ leads to MMSE
log-spectral amplitude estimator in {\cite{ephraim1985speech}}. More complicated forms of $f(a)$ and $g(a,b,c)$ can be chosen, for example, many perceptually-weighted
error criteria can be included, which can be referred to {\cite{hu2003perceptually}}.


For supervised approaches, the square error in the subband is often defined as the loss function in the fullband, which is
\begin{equation}
\label{eq4}
\begin{array}{l}
\mathcal{J}_x  = \sum\limits_{k \in \mathcal{K}} {\sum\limits_{l \in \mathcal{L}} {J_x \left[ {M_l \left( k \right)} \right]} }.
 \end{array}
\end{equation}
One can get that, when $f(a) = \log(|a|)$ and $g(a,b,c) = \log(|(a+b)c|)$, $\mathop {\min }\limits_{M_l \left( k \right)} \left\{\mathcal{J}_x \right\}$ is to minimize the MSE of log-spectral amplitude between the clean speech
and the estimated speech, which is the training target in {\cite{xu2013experimental}}.


Note that (\ref{eq3}) and (\ref{eq4}) are quite similar and the most obvious difference between them
is that $J_x \left[ {M_l\left( {k} \right)} \right]$ is the subband square error, while $\mathcal{J}_x$
is the fullband square error. The other difference is that the nonlinear spectral gain can be derived theoretically by
minimizing $E\left\{J_x \left[ {M_l\left( {k} \right)} \right]\right\}$ when the probability density function (p.d.f.) of the
speech and that of the noise are both given, while it is difficult to derive the nonlinear spectral gain by minimizing $\mathcal{J}_x$,
where this gain can often be mapped from the input noisy features after training the supervised machine learning model. In all, it seems that all subband square error functions can be generalized to the fullband ones as supervised training targets.
\vspace{-0.3cm}
\section{Proposed Algorithm}
\vspace{-0.1cm}
Only using MMSE as a criterion, it is difficult to make a balance between speech distortion and noise reduction. This section derives a more generalized fullband loss function.

\subsection{Trade-off Criterion in Subband}
In traditional speech enhancement approaches, speech distortion and noise reduction in the subband can be considered separately. The subband square error of the speech and the subband residual noise
can be, respectively, given by
\begin{equation}
\label{eq5}
J_s \left[ {M_l \left( k \right)} \right] = {\left| {f\left( {S_l \left( k \right)} \right) - g\left( {S_l \left( k \right),D_l \left( k \right),M_l \left( k \right)} \right)} \right|^2 },
\end{equation}
and
\begin{equation}
\label{eq6}
J_d \left[ {M_l \left( k \right)} \right] = {\left| {h\left( {S_l \left( k \right),D_l \left( k \right),M_l \left( k \right)} \right)}\right| }^2,
\end{equation}
where $h(a,b,c)$ is a function with three variables $a$, $b$, and $c$. When $f(a) = |a|$, $g(a,b,c) = |ac|$, and $h(a,b,c) = |bc|$, $E\left\{ J_s \left[ {M_l \left( k \right)} \right]\right\}$
and $E\left\{ J_d \left[ {M_l \left( k \right)} \right]\right\}$ become the MSE of the
speech magnitude and the residual noise power in the subband, respectively, which are identical with \cite[(8.31) and (8.32)]{benesty2009noise}.


By minimizing the subband MSE of the speech with a residual noise control, an optimization problem
can be given to derive the nonlinear spectral gain, which is given by
\begin{equation}
\label{eq7}
\begin{array}{l}
\mathop {\min}\limits_{M_l \left( k \right)} E \left\{ {J_s \left[ {M_l \left( k \right)} \right]} \right\},\\
\;\;\;\;\;\;{\emph s.t.}\;\;E\left\{J_d \left[ {M_l \left( k \right)} \right]\right\} = {\left| {\mathchar'26\mkern-10mu\lambda \left( {\beta _l \left( k \right),D_l \left( k \right)} \right)} \right|^2 },
\end{array}
\end{equation}
where ${\mathchar'26\mkern-10mu\lambda \left( {\beta ,b} \right)}$ is a function of two variables $\beta$ and $b$. $\beta _l \left( k \right) \in [0\;1]$ could be both a frequency and frame-dependent factor that can be introduced to control the residual noise flexibly.

The optimal spectral gain in (\ref{eq7}) can be solved theoretically by the Lagrange multiplier method, which is
\begin{equation}
\label{eq8}
\mathop {\min }\limits_{M_l \left( k \right)} \left\{ \begin{array}{l}
 E\left\{J_s \left[ {M_l \left( k \right)} \right]\right\} + \mu E\left\{ J_d \left[ {M_l \left( k \right)} \right]\right\} \\
  \;\;\;\;\;\;\;\;\;\;\;\;\;\;\;\;\;\;\;\;\;\;\;\;\;\;\; - \mu  {\left| {\mathchar'26\mkern-10mu\lambda \left( {\beta _l \left( k \right),D_l \left( k \right)} \right)} \right|^2 }  \\
 \end{array} \right\},
\end{equation}
where $\mu \ge 0$ is a Lagrange multiplier. When $f(a) = |a|$, $g(a,b,c) = |ac|$, $h(a,b,c) = |bc|$, and $\left|{\mathchar'26\mkern-10mu\lambda \left( {\beta ,b} \right)}\right| = \beta E\{|b|^2\}$,
the optimal spectral gain can be derived from ({\ref{eq8}}) and the constraint in (\ref{eq7}), which can be given by
\begin{equation}
\label{eq9}
M_l \left( k \right) = {{\xi _l \left( k \right)} \mathord{\left/
 {\vphantom {{\xi _l \left( k \right)} {\left( {\xi _l \left( k \right) + \mu _l \left( k \right)} \right)}}} \right.
 \kern-\nulldelimiterspace} {\left( {\xi _l \left( k \right) + \mu _l \left( k \right)} \right)}},
\end{equation}
where $\xi_l\left( k \right) = E\{|S_l \left( k \right)|^2\}/E\{|D_l \left( k \right)|^2\}$ is the \emph{a priori} SNR. It is not always possible to derive $M_l \left( k \right)$ mathematically, especially when
$f(a)$, $g(a,b,c)$, $h(a,b,c)$, and ${\mathchar'26\mkern-10mu\lambda \left( {\beta ,b} \right)}$ have very complicated expressions. Moreover, it is uneasy to accurately estimate the noise power spectral density in non-stationary noise environments~{\cite{martin2001noise,cohen2003noise, gerkmann2011unbiased}}.
However, it seems that this optimization can be easily solved by supervised approaches. To transfer this problem, we need to
define the fullband square error of the speech and the fullband residual noise power to derive the loss function for supervised approaches.
\textbf{\vspace{-0.3cm}}
\subsection{Trade-off Criterion in Fullband}
The fullband MSE of the speech and the fullband residual noise can be, respectively, given by
\begin{equation}
\label{eq10}
\mathcal{J}_s  = \sum\limits_{k = \mathcal{K}} {\sum\limits_{l = \mathcal{L}} {J_s \left[ {M_l \left( k \right)} \right]} },
\end{equation}
and
\begin{equation}
\label{eq11}
\mathcal{J}_d  = \sum\limits_{k = \mathcal{K}} {\sum\limits_{l = \mathcal{L}} {J_d \left[ {M_l \left( k \right)} \right]} }.
\end{equation}

The loss function without any constraints can be given by
\begin{equation}
\label{eq12}
\mathcal{J}_x  = \mathcal{J}_s  + \mu \mathcal{J}_d,
\end{equation}
where (\ref{eq12}) is the same as the newly proposed components loss function as given in~{\cite{xu2019components}}.

The loss function with residual noise control is
\begin{equation}
\label{eq13}
\mathcal{J}_x  = \mathcal{J}_s  + \mu \mathcal{J}_d^{\rm con},
\end{equation}
where
\begin{equation}
\mathcal{J}_d^{\rm con}  = \sum\limits_{k = \mathcal{K}} {\sum\limits_{l = \mathcal{L}} {\left| {J_d \left[ {M_l \left( k \right)} \right] -  {\left| {\mathchar'26\mkern-10mu\lambda \left( {\beta_l \left( k \right),D_l \left( k \right)} \right)} \right|^2 } } \right|} }.\nonumber
\end{equation}

It is obvious that (\ref{eq13}) is a generalization of (\ref{eq12}), where (\ref{eq13}) reduces to (\ref{eq12}) when
${ {\left| {\mathchar'26\mkern-10mu\lambda \left( {\beta_l \left( k \right),D_l \left( k \right)} \right)} \right|^2 } \equiv 0}$. One can observe that $\beta_l \left( k \right)$
is both frequency and frame-dependent, so it can control the residual noise in each time-frequency bin.
\vspace{-0.3cm}
\subsection{A Generalized Loss Function}
We further generalize the subband square error in (\ref{eq5}) and (\ref{eq6}), the square is substituted by a variable $\gamma \ge 0$ and an additional
 variable $\alpha$ is also introduced on the spectra, then (\ref{eq5}) and (\ref{eq6}) can be, respectively, given by
\begin{equation}
\label{eq14}
J_s^{\gamma,\alpha}  \left[ {M_l \left( k \right)} \right] =  {\left| {f\left( {S^\alpha_l \left( k \right)} \right) - g\left( {S^\alpha_l \left( k \right),X^\alpha_l \left( k \right),M^\alpha_l \left( k \right)} \right)} \right|^\gamma  } ,
\end{equation}
and
\begin{equation}
\label{eq15}
J_d^{\gamma,\alpha}  \left[ {M_l \left( k \right)} \right] =  {\left| {h\left( {S^\alpha_l \left( k \right),D^\alpha_l \left( k \right),M^\alpha_l \left( k \right)} \right)} \right|^\gamma  }.
\end{equation}

Analogously, with the residual noise control, the optimal problem in the subband becomes
\begin{equation}
\label{eq16}
\begin{array}{l}
\mathop {\min }\limits_{M_l \left( k \right)} E \left\{ {J_s^{\gamma,\alpha} \left[ {M_l \left( k \right)} \right]} \right\},\\
\;\;\;\;\;\;{\rm \emph{s.t.}}\;\;\;E\left\{J_d^{\gamma,\alpha} \left[ {M_l \left( k \right)} \right]\right\} = {\left| {\mathchar'26\mkern-10mu\lambda \left( {\beta^\alpha_l \left( k \right),D^\alpha_l \left( k \right)} \right)} \right|^{\gamma} }.
\end{array}
\end{equation}

By setting $f(a) = |a|$, $g(a,b,c) = |ac|$, $h(a,b,c) = |bc|$, and ${\mathchar'26\mkern-10mu\lambda \left( {\beta ,b} \right)} = (\beta |b|)$, one can derive a generalized
gain function with the Lagrange multiplier method, which is
\begin{equation}
\label{eq:ParametricWF}
M_l \left( k \right) = \left( {\frac{{\left( {\xi _l \left( k \right)} \right)^{c_1 } }}{{\left(\mu _l \left( k \right)\right)^{\left(2 c_1c_2-1\right)} + \left( {\xi _l \left( k \right)} \right)^{c_1 } }}} \right)^{c_2 },
\end{equation}
where $c_1 = \alpha\gamma/(2\gamma-2)$ and $c_2 = 1/\alpha$, where (\ref{eq:ParametricWF}) is identical to \cite[(6)]{inoue2011theoretical}. Note that \cite[(6)]{inoue2011theoretical} is
given intuitively without theoretical derivation. When $\gamma=2$ and $\alpha=1$, (\ref{eq:ParametricWF}) reduces to (\ref{eq9}). When $\gamma=2$, one can get $M_l \left( k \right) = \left( {{{\left( {\xi _l \left( k \right)} \right)^\alpha  } \mathord{\left/
 {\vphantom {{\left( {\xi _l \left( k \right)} \right)^\alpha  } {\left( {\mu _l \left( k \right) + \left( {\xi _l \left( k \right)} \right)^\alpha  } \right)}}} \right.
 \kern-\nulldelimiterspace} {\left( {\mu _l \left( k \right) + \left( {\xi _l \left( k \right)} \right)^\alpha  } \right)}}} \right)^{1/\alpha }$, which has already been derived and presented in (\cite[(22)]{inoue2011theoretical}).

Similarly, the generalized loss function for supervised approaches can be given by
\begin{equation}
\label{eq17}
\mathcal{J}_x^{\gamma,\alpha}  = \mathcal{J}_s^{\gamma,\alpha}  + \mu \mathcal{J}_d^{{\gamma,\alpha},\rm con},
\end{equation}
where the first item $\mathcal{J}_s^{\gamma,\alpha}  = \sum\limits_{k = \mathcal{K}} {\sum\limits_{l = \mathcal{L}} {J_s^{\gamma,\alpha} \left[ {M_l \left( k \right)} \right]} }$ relates to the fullband speech distortion and the second item
$\mathcal{J}_d^{{\gamma,\alpha}, \rm con}  = \sum\limits_{k = \mathcal{K}} {\sum\limits_{l = \mathcal{L}} {\left| {J_d^{\gamma,\alpha} \left[ {M_l \left( k \right)} \right] - {\left| {\mathchar'26\mkern-10mu\lambda \left( {\beta ^\alpha_l \left( k \right),D^\alpha_l \left( k \right)} \right)} \right|^\gamma } } \right|} }$
is introduced to control the residual noise.

Eq. (\ref{eq17}) is a generalized loss function that includes (\ref{eq12}) and (\ref{eq13}). This is because (\ref{eq17}) reduces to (\ref{eq13}) when $\gamma=2, \alpha=1$ and it can further reduces to (\ref{eq12})
by setting ${{\left| {\mathchar'26\mkern-10mu\lambda \left( {\beta^\alpha _l \left( k \right),D^\alpha_l \left( k \right)} \right)} \right|^\gamma } \equiv 0}$. It is interesting to see that (\ref{eq3})
also can be separated into two components, where one is the MSE of the speech and the other is related to the residual noise. When $f(a) = a$ and $g(a,b,c) = (a+b)c$, we have
\begin{equation}
\label{eq18}
E\left\{ {J_x \left( {M_l \left( k \right)} \right)} \right\} = E\left\{ {J_s \left( {M_l \left( k \right)} \right)} \right\} + E\left\{ {J_d \left( {M_l \left( k \right)} \right)} \right\},
\end{equation}
where $E\left\{J_s \left( {M_l \left( k \right)} \right)\right\} = \left| {1 - M_l \left( k \right)} \right|^2 E\left\{ {\left| {S_l \left( k \right)} \right|^2 } \right\}$ relates to the power of speech distortion and $E\left\{ {J_d \left( {M_l \left( k \right)} \right)} \right\} = \left| {M_l \left( k \right)} \right|^2 E\left\{ {\left| {D_l \left( k \right)} \right|^2 } \right\}$ relates to the power of residual noise. $E\left\{J_x \left[ {M_l \left( k \right)} \right]\right\}$ is a combination of speech distortion and residual noise, so the fullband MSE loss function of a complex spectrum is also a special case of the generalized loss function in (\ref{eq17}). If $f(a) = |a|$ and $g(a,b,c) = |(a+b)c|$ are chosen, the decomposition of $E\left\{ {J_x \left( {M_l \left( k \right)} \right)} \right\}$ is more complicated than (\ref{eq18}), which will not be further discussed for limited space.

In this letter, we emphasize the importance of introducing the residual noise control. $f(a) = |a|$, $g(a,b,c) = |ac|$, $h(a,b,c) = |bc|$, and ${\mathchar'26\mkern-10mu\lambda \left( {\beta ,b} \right)} = (\beta |b|)$ are applied, although more complicated expressions can be chosen when taking the perceptual quality into account. Accordingly, we have
\begin{equation}
\label{eq19}
\mathcal{J}_s^{\gamma,\alpha}  = \sum\limits_{l = \mathcal{L}} {\sum\limits_{k = \mathcal{K}} {\left| {\left( {1 - M^\alpha_l \left( k \right)} \right)S^\alpha_l \left( k \right)} \right|^\gamma  } },
\end{equation}
and
\begin{small}
\begin{equation}
\label{eq20}
\mathcal{J}_d^{{\gamma,\alpha} ,con}  = \sum\limits_{l = \mathcal{L}} {\sum\limits_{k = \mathcal{K}} {\left| {\left| {M_l \left( k \right)D_l \left( k \right)} \right|^{\alpha\gamma} - \left| {\beta _l \left( k \right)D_l \left( k \right)} \right|^{\alpha\gamma}} \right|} },
\end{equation}
\end{small}
where $\alpha$ will be set to a constant value and $\beta _l \left( k \right)$ is a constant value over frequency for simplicity, that is to say, $\beta _l \left( k \right) \equiv \beta_0$ and $\alpha=1$ are used in the following.
We only study the impact of $\beta_0$, $\mu$, and $\gamma$ on supervised approaches.

\section{Experimental Setup}
\subsection{Dataset}
Experiments are conducted with TIMIT corpus, where 1000 and 200 utterances are randomly chosen as the training and the evaluation datasets, respectively. 125 types of environment noises~{\cite{xu2013experimental, duan2012speech}} are used for generating noisy utterances under different SNR levels ranging from -5dB to 15dB with the interval 5dB. For model test, additional 10 male and 10 female utterances are chosen to mix with unseen noise signals taken from the NOISEX92~{\cite{varga1993assessment}} with SNR ranging from -5dB to 10dB with the interval 5dB.

\subsection{Network Architecture}
\begin{figure}[t]
	\setlength{\abovecaptionskip}{0.235cm}
	\setlength{\belowcaptionskip}{-0.1cm}
	\centering
	\centerline{\includegraphics[width= \columnwidth]{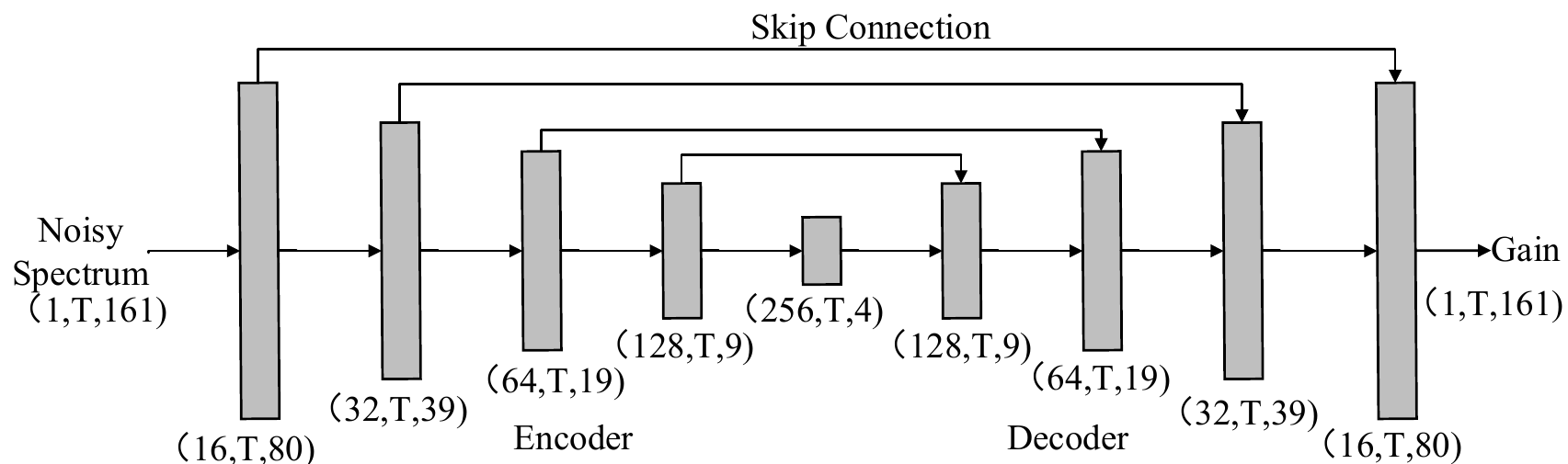}}
	\caption{The network architecture adopted in this study. Input is the noisy magnitude spectra and output is the estimated gain functions.}
	\label{fig:u-net}
	\vspace{-0.3cm}
\end{figure}

\vspace{-0cm}
U-Net is chosen as the network in this letter, which has been widely adopted for speech separation task~{\cite{kolbaek2019loss}}. As shown in Fig.~{\ref{fig:u-net}}, the network consists of the convolutional encoder and decoder, both of which are comprised of five convolutional blocks where the 2-D convolution layer is adopted, followed by batch normalization (BN) and exponential linear unit (ELU). Skip connection is introduced to compensate for the information loss during features compression process. Note that the mapping target is the gain function and the sigmoid function is adopted to make sure that the output ranges from 0 to 1. Causal mechanism is introduced to achieve real-time processing, where only the past frames are involved in the convolution calculation. The tensor output size of each layer is given with $(Channels, TimeStep, Feat)$ format, which is shown in Fig.~{\ref{fig:u-net}}.
\vspace{-0.3cm}
\subsection{Loss Functions and Training Models}
This letter chooses three loss functions including MSE in~(\ref{eq4}), Time-MSE-based loss (TMSE)~{\cite{kolbaek2019loss}} and recently proposed SI-SDR-based loss~{\cite{kolbaek2019loss}} as baselines. AS T-F domain-based network is used, an additional fixed iSTFT-like layer is needed to transform the estimated T-F spectrum back into time domain for TMSE- and SI-SDR-based loss~{\cite{wichern2018phase}}. They compare with the proposed generalized loss function given in (\ref{eq17}) with (\ref{eq19}) and (\ref{eq20}). All the models are trained with stochastic gradient descent (SGD) optimized by Adam~{\cite{kingma2014adam}}.



\vspace{-0.1cm}
\section{Results and Analysis}
\vspace{-0.1cm}
\subsection{Objective Evaluation}
\begin{figure}[htbp]
	\setlength{\abovecaptionskip}{0.235cm}
	\setlength{\belowcaptionskip}{-0.1cm}
	\centering
	\centerline{\includegraphics[width= \columnwidth]{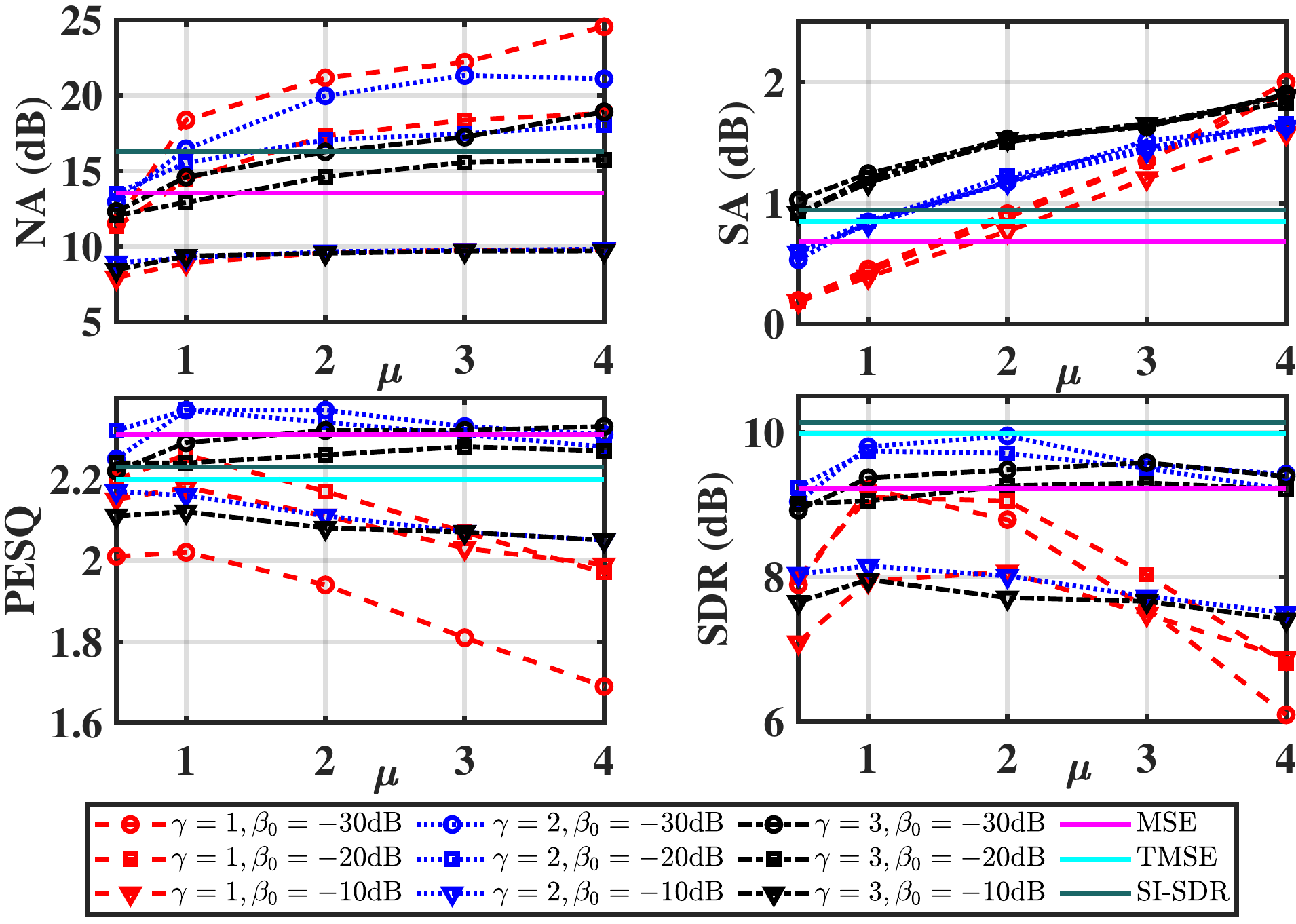}}
	\caption{Test results in terms of NA, SA, PESQ and SDR, where the averaged PESQ score of the noisy signals is 1.80 and its averaged SDR is 2.51{\rm dB}.}
	\label{fig:na_sa_pesq}
	\vspace{-0.3cm}
\end{figure}

This letter uses four objective measurements including noise attenuation (NA)~{\cite{gustafsson1998novel}}, speech attenuation (SA)~{\cite{gustafsson1998novel}}, PESQ~{\cite{recommendation2001perceptual}}, and SDR~{\cite{vincent2006performance}}. The testing results w.r.t. $\gamma, {\beta_0}$ and $\mu$ are shown in Fig.~\ref{fig:na_sa_pesq}, where $\gamma=1,2,3$, $\beta_0=-10{\rm dB},-20{\rm dB},-30{\rm dB}$ and $\mu=0.5,1,2,3,4$ are considered. The test results of three baselines are also presented as comparison. From this figure, one can observe the following phenomena. First, the increase of ${\beta_0}$ will decrease NA. This is because the residual noise control mechanism is introduced for optimization, which means, during the training process, the residual noise in the estimated spectra will gradually get close to the preset residual noise threshold. As a consequence, the characteristic of the residual noise is expected to be effectively preserved, which will be further confirmed by subjective listening tests in the following. Second, the increase of $\mu$ is beneficial to noise suppression and meanwhile introducing more speech distortion. As generalized loss can be viewed as the joint optimization of both speech distortion and noise reduction, a larger $\mu$ leads to smaller gain values, as (\ref{eq:ParametricWF}) states, where on the one hand more interference is suppressed and on the other hand, more speech components are inevitably abandoned. Third, the increase of $\gamma$ has a negative influence on NA and SD. Finally, among various parameter configurations, (2, -30{\rm dB}, 0.5), (2, -30{\rm dB}, 1) and (2, -20{\rm dB}, 1) can be chosen. This is because relatively better performance can be obtained for all the four objective metrics. One can observe that the three competing loss functions can get better performance in some objective metrics, while they may suffer much worse performance in others. For example, SI-SDR and TMSE have larger values of SDR, while their PESQ scores are even lower than the MSE, which is consistent with the study in {\cite{le2019sdr}}.

\subsection{Subjective Evaluation}
To evaluate speech quality of the proposed generalized loss (GL) function, a subjective evaluation test is conducted among GL and baselines, where we follow the subjective testing procedures of~{\cite{breithaupt2007cepstral}}. In this comparison, we choose the parameter configuration (2, -20{\rm dB}, 1) for the propose GL function. The experiment is conducted in a standard listening room, where 10 listeners participate. The listening material consists of 20 utterances, each of which includes one male and female utterance selected from TIMIT corpus and is mixed with one of five noises including aircraft, babble, bus, cafeteria, and car. Four SNR conditions are selected for mixing, i.e. -5{\rm dB}, 0{\rm dB}, 5{\rm dB}, 10{\rm dB}. Speech pause of 3{\rm s} duration is specifically inserted before each utterance. Then, the duration of each listening utterance is about 13{\rm s}. Each listener needs to write down the utterance index that they prefer considering both noise naturalness and speech quality. The same as ~{\cite{breithaupt2007cepstral}}, "Equal" option is also provided if no subjective preference can be given. To avoid inertia, the utterance index in each pair is shuffled. The averaged subjective results are presented in Table.{~\ref{tab1}}. From this table, one can observe that the proposed GL function with residual noise control achieves better performance in subjective testing, which can be explained as the proposed GL method can effectively recover speech components while preserving the characteristic of background noise to some extent compared with all the baselines.
{
	\renewcommand\arraystretch{1.2}
	\setlength{\tabcolsep}{18pt}
	\begin{table}[h]
		\tiny
		\footnotesize
		\caption{Results of subjective listening test. The numbers indicate the percentage of votes in favor of one approach. The choice "Equal" means no subjective difference.}
		\centering
		\begin{center}
			\begin{tabular}{c|c|c|c}
				\hline\hline
				\textbf{Methods} & \textbf{GL} & \textbf{MSE} & \textbf{Equal} \\
				\hline
				\textbf{Preference}  & 70.0\%  & 22.0\% & 8.0\% \\
				\hline
				\textbf{Methods} & \textbf{GL} &\textbf{TMSE} & \textbf{Equal} \\
				\hline
				 \textbf{Preference} & 66.5\% & 22.0\% & 12.5\% \\
				\hline
				\textbf{Methods} & \textbf{GL} & \textbf{SI-SDR} & \textbf{Equal} \\
				\hline
				\textbf{Preference} & 70.5\% & 23.5\% & 6.0\% \\
				\hline\hline
			\end{tabular}
			\label{tab1}
		\end{center}
	\end{table}
}

\section{Conclusion}
This letter derives a generalized loss function which can easily make a balance between noise attenuation and speech distortion with multiple manual parameters. In addition, MSE and other typical loss functions are revealed to be special cases. Both objective and subjective tests are conducted to show that it is important to control the residual noise for supervised speech enhancement approaches, where the residual noise becomes much more natural than before. Further work could concentrate on studying a combination of the residual noise control scheme with objective metrics-based loss functions to improve the naturalness of the residual noise.


%

%

%
%

\ifCLASSOPTIONcaptionsoff
  \newpage
\fi

\end{document}